\documentclass[aps,twocolumn]{revtex4}%
\usepackage{amsfonts}
\usepackage{amsmath}
\usepackage{amssymb}
\usepackage{graphicx}%
\begin{document}
\title{Protocol for universal gates in optimally biased superconducting qubits}
\author{Chad Rigetti and Michel Devoret}
\affiliation{Department of Applied Physics, Yale University, New Haven, Connecticut 06520-8284}

\begin{abstract}
We present a new experimental protocol for performing universal gates in a
register of superconducting qubits coupled by fixed on-chip linear reactances.
The qubits have fixed, detuned Larmor frequencies and can remain, during the
entire gate operation, biased at their optimal working point where decoherence
due to fluctuations in control parameters is suppressed to first order.
Two-qubit gates are performed by simultaneously irradiating two qubits at
their respective Larmor frequencies with appropriate amplitude and phase,
while one-qubit gates are performed by the usual single-qubit irradiation pulses.

\end{abstract}
\volumeyear{year}
\volumenumber{number}
\issuenumber{number}
\eid{identifier}
\received[Received text]{date}

\revised[Revised text]{date}

\accepted[Accepted text]{date}

\published[Published text]{date}

\startpage{1}
\endpage{2}
\maketitle

Single quantum bits displaying coherence in the time domain have
now been implemented in various superconducting integrated
electrical circuits\cite{One Qubit}. Microwave
spectroscopy\cite{Majer Berkley}, coherent temporal
oscillations\cite{Pashkin}, and a conditional gate
operation\cite{Yamamoto} have been reported in experiments on
pairs of capacitively coupled qubits. In all these
implementations, decoherence is by far the largest obstacle to be
overcome for applications to quantum information processing. Yet,
as Vion \textit{et al.} have demonstrated, appropriate symmetries
in circuit architecture and bias conditions can be exploited for
suppressing to first order decoherence due to fluctuations in
control parameters.

The schemes for performing two-qubit gates proposed so far require dynamical
tuning of either the qubit transition frequencies\cite{Majer Berkley} or of
the subcircuit controlling the qubit-qubit interaction\cite{BlaisAverinWhaley}%
. The former typically requires dc pulses that move the qubits away from their
optimal bias points for coherence, while the latter requires additional
control lines and non-linear elements that inevitably introduce additional
couplings to uncontrolled degrees of freedom in the environment. In this
Letter, we present a novel scheme which minimizes decoherence by maintaining
both qubits at their optimal bias points, and by employing only noise-free
fixed linear coupling reactances. Furthermore, this scheme takes advantage of
the spread in circuit parameters occuring naturally in fabrication, instead of
suffering from it.

Our strategy consists of constructing a circuit with fixed, detuned Larmor
frequencies and fixed coupling strengths---a sort of \textquotedblleft
artificial molecule\textquotedblright---and realizing gates with protocols
inspired by those of nuclear magnetic resonance(NMR) quantum
computation\cite{NMR1}. The essential difference between our \textquotedblleft
molecules\textquotedblright\ and those used in NMR resides in the form of the
qubit--qubit couplings and the way they are exploited. In NMR, the secular
terms in the coupling Hamiltonian (i.e. those that commute with the Zeeman
Hamiltonian and thus act to first order) dominate the spin-spin interaction.
Two-qubit gates are realized as the spins precess freely, while refocusing
pulses are applied in order to \textquotedblleft do nothing\textquotedblright.
In our scheme, the coupling is purely \textit{non}-secular, and has no effect
to first order. So unlike in NMR, we must construct pulses to enhance the
second-order effect of the coupling. We refer to this strategy with the
(NMR-style!) nickname: FLICFORQ, for Fixed LInear Couplings between Fixed
Off-Resonant Qubits.

The superconducting register we have in mind may consist of charge qubits
(controlled via charges on gate capacitors) interacting through on-chip
capacitors or of flux qubits (controlled via fluxes through superconducting
loops) interacting through mutual inductances. We focus for the moment on
two-qubit registers (Fig. 1), the simplest that allow the realization of a
universal set of quantum gates, leaving the extension to larger systems to the
discussion. The optimal bias conditions for the circuits shown are $N_{1}%
^{g}=N_{2}^{g}=1/2$ for charge qubits, where $N^{g}=C^{g}U/2e$ is the
dimensionless gate charge, or $N_{1}^{\phi}=N_{2}^{\phi}=1/2$ for flux qubits,
where $N^{\phi}=\Phi^{ext}/\Phi_{o}$ is the flux frustration. Under these
conditions, the systems become immune, to first order, to variations in the
control parameters, such as $1/f$ charge noise in the Josephson junctions or
substrate or noise due to the motion of trapped flux\cite{Dev/Mart}.
\begin{figure}[ptbh]
\includegraphics[width=3.1in]{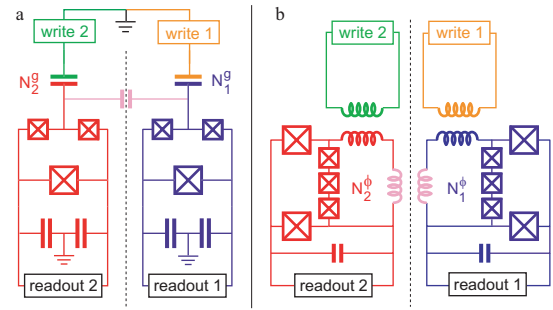}
\caption{Superconducting two-qubit circuits for performing universal quantum
gates at optimal bias point. (a) Charge qubits (Saclay style) coupled by a
capacitor. (b) Flux qubits (Delft style) coupled by a mutual inductance.}%
\end{figure}

At optimal bias, these two-qubit systems are described by the reduced
Hamiltonian
\begin{align*}
\mathcal{H}/\hbar &  =\frac{1}{2}[\omega_{1}^{z}\sigma_{1}^{z}+2[\omega
_{1}^{x}(t)\cos(\omega_{1}^{rf}t)+\omega_{1}^{y}(t)\sin(\omega_{1}%
^{rf}t)]\sigma_{1}^{x}\\
+  &  \omega_{2}^{z}\sigma_{2}^{z}+2[\omega_{1}^{x}(t)\cos(\omega_{2}%
^{rf}t)+\omega_{2}^{y}(t)\sin(\omega_{2}^{rf}t)]\sigma_{2}^{x}\\
&  +\omega^{xx}\sigma_{1}^{x}\sigma_{2}^{x}],
\end{align*}
where $\omega_{1}^{z}/2\pi$ ($\omega_{2}^{z}/2\pi$) is the Larmor frequency of
qubit 1(2); $\omega_{1}^{rf}/2\pi$ ($\omega_{2}^{rf}/2\pi$) is the frequency
of the signal applied to the \textquotedblleft write\textquotedblright\ port
of qubit 1(2); $\omega_{1}^{x}$ ($\omega_{2}^{x}$) and $\omega_{1}^{y}$
($\omega_{2}^{y}$) are the amplitudes of the in-phase and quadrature
components of the applied signals, respectively, and, when divided by $2\pi$,
are directly interpretable as Rabi frequencies; and $\omega^{xx}%
/2\pi=(t^{swap})^{-1}$ is the \textquotedblleft swap\textquotedblright%
\ frequency (if only the $\sigma_{1}^{x}\sigma_{2}^{x}$ term were present in
$\mathcal{H}$, the time needed to go from a product state to a maximally
entangled state would be $t^{swap}/4$). The Larmor frequencies are detuned
from one another, as occurs naturally during fabrication, by $\delta
=\omega_{1}^{z}-\omega_{2}^{z}$, and remain fixed throughout. The swap
frequency is fixed at the time of circuit fabrication and should satisfy
$\omega^{xx}\ll\delta$ to avoid significant entanglement of the qubits in the
absence of external signals. For concreteness, we consider the case
$\omega^{xx}=0.1\delta=0.01\omega_{o}$, where $\omega_{o}=(\omega_{1}%
^{z}+\omega_{2}^{z})/2$, but our results do not depend critically on these
values. For simplicity, we limit ourselves in this paper to resonant RF pulses
constrained to obey $\omega_{1}^{rf}=\omega_{1}^{z}$ and $\omega_{2}%
^{rf}=\omega_{2}^{z}$ and we perform all possible gates by playing with only
four external knobs $\omega_{1}^{x},\omega_{1}^{y},\omega_{2}^{x}$ and
$\omega_{2}^{y}$. The difference between $\omega_{1}^{rf}$ and $\omega
_{2}^{rf}$ suppresses cross-talk during gate operations, a crucial practical
advantage of FLICFORQ.

The mechanism allowing the very weak interqubit coupling $\omega^{xx}$ to
produce maximally entangled two-qubit states is easily understood in the
dressed atom picture of quantum optics\cite{Cohen-Tannoudji,note on classical
comp}. When the RF fields and qubits are uncoupled, each qubit + field system
has an infinite discrete ladder of doubly-degenerate energy levels, labelled
by the qubit state $|1\rangle$ or $|0\rangle$ and the photon number
$|N\rangle$, and separated by $\omega^{rf}$ (Fig. 2, outer levels). Taking the
qubit--field coupling into account lifts the degeneracy, causing the two
states in each manifold to be split symmetrically by the field strength
$\omega_{1}^{y}$ (Fig 2, inner levels). The two dressed qubits may then absorb
and emit energy at frequencies $\omega_{1}^{z}\pm\omega_{1}^{y}$ and
$\omega_{2}^{z}\pm\omega_{2}^{y}$, respectively. The result of the irradiation
is thus to split the single-mode qubit line into two sidebands at these
frequencies. Choosing the RF amplitudes $\omega_{1,2}^{y}=\delta/2$ causes the
upper sideband of qubit 1 to overlap the lower sideband of qubit 2, allowing
the qubits to exchange photons of energy $\hbar(\omega_{1}^{z}-\omega_{1}%
^{y})=\hbar(\omega_{2}^{z}+\omega_{2}^{y})$ through the coupling
reactance.\begin{figure}[pb]
\begin{center}
\includegraphics[width=3.4in
]{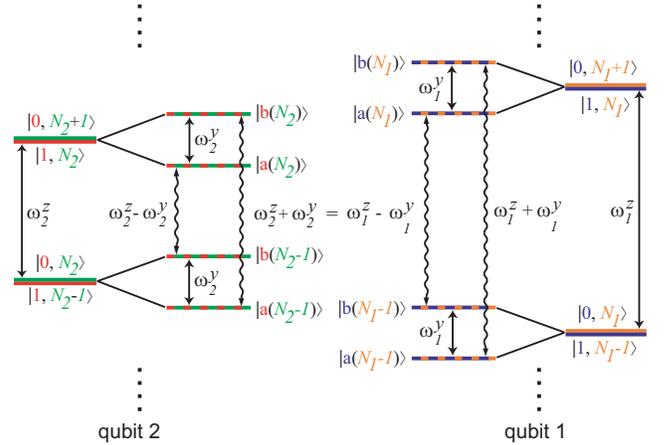}
\end{center}
\caption{Energy levels of qubit + RF photons systems with (inner levels) and
without (outer levels) qubit-photon coupling. \emph{Outer}: systems have an
infinite ladder of doubly-degenerate levels corresponding to products of a
photon number state (green, orange) and a qubit state (red, blue).
\emph{Inner}: Photon--qubit coupling lifts degeneracy in each manifold by Rabi
frequency $\omega^{y}$. Transitions between adjacent manifolds (wavy arrows)
correspond to absorption/emission of a photon from dressed qubit system.
Transition of qubit 1 at $\omega_{1}^{z}-\omega_{1}^{y}$ and qubit 2 at
$\omega_{2}^{z}+\omega_{2}^{y}$ coincide when $\omega_{1}^{y}=\omega_{2}%
^{y}=\delta/2$, putting qubits on speaking terms.}%
\end{figure}

The simplest protocol for generating entangled states in this system is to
simultaneously irradiate each qubit with a signal of amplitude $\delta
/2$\cite{comment 1}. If we choose $\omega_{1,2}^{y}=\delta/2$ and
$\omega_{1,2}^{x}=0$, and initialize the state to $\mathbf{\rho}%
_{in}=|00\rangle\langle00|$, the qubits will become maximally entangled after
a pulse time $4\pi/\omega^{xx}=2t^{swap}$. If the RF is then switched off, the
system will remain in an entangled state until it is measured in a local basis
or it undergoes decoherence or relaxation. Note that this scheme allows us to
produce entanglement on demand without dc excursions from the optimal bias
point of either qubit.

The rotation realized when the system is irradiated in this manner, which we
call \textquotedblleft\textsc{D}\textquotedblright, is not a pure\ $\sigma
_{1}^{x}\sigma_{2}^{x}$ rotation, but is rather a product of two commuting
$\pi/2$ rotations:%
\begin{align}
\text{\textsc{D}}  &  =(\mathbf{1}-i\sigma_{1}^{z}\sigma_{2}^{z}%
)(\mathbf{1}+i\sigma_{1}^{x}\sigma_{2}^{x})/2\\
&  =(Z_{1}Z_{2})^{-1/2}(X_{1}X_{2})^{1/2},\nonumber
\end{align}
where we have used the rotation operator notation in which $X_{1}=i\sigma
_{1}^{x},$ $X_{1}X_{2}=i\sigma_{1}^{x}\sigma_{2}^{x}$, \textit{etc}. This
rotation maps the computational basis states to the Bell states with a
relative phase, e.g. \textsc{D}: $|00\rangle\rightarrow(|00\rangle
+i|11\rangle)/\sqrt{2}$\cite{Rig/Dev GDQI}, and can therefore be used to
generate and study maximally entangled states. However, it is easy to verify
that \textsc{D}$^{2}=-1$, indicating that \textsc{D} is a $\pi$ rotation, and
therefore is not universal\cite{Nielsen and Chuang}.

We propose to circumvent this problem by nulling out the $\sigma_{1}^{z}%
\sigma_{2}^{z}$ factor of \textsc{D}. This is done by flipping the sign of the
RF signal amplitude on one of the two qubits midway through the pulse. With
this \textquotedblleft refocusing flip\textquotedblright\ the unwanted
$\sigma_{1}^{z}\sigma_{2}^{z}$ rotation taking place during the first half of
the pulse will be fully undone during the second half. This technique
resembles the refocusing schemes used in NMR\cite{NMR2}, though here we are
modifying the pulse shape rather than performing additional $\pi$ rotations.

Implementing the refocusing flip leads to the pure $\pi/2$ rotation
$(X_{1}X_{2})^{1/2}=(\mathbf{1}+i\sigma_{1}^{x}\sigma_{2}^{x})/\sqrt{2}$. This
rotation, when augmented by one-qubit $\pi/2$ rotations, is known to generate
the two-qubit Clifford group $\mathcal{C}_{2}$\cite{Rig/Dev GDQI}. So along
with all one-qubit unitaries, $(X_{1}X_{2})^{1/2}$ therefore constitutes a
universal set of rotations.

We can thus turn to the construction of a protocol to perform
$U^{\text{\textsc{CNOT}}}$, the rotation corresponding to the standard
two-qubit logical gate \textsc{CNOT}. We first decompose
$U^{\text{\textsc{CNOT}}}$ into a sequence of rotations that draws only on
$(X_{1}X_{2})^{1/2}$ and one-qubit $\pi/2$ rotations. We use the sequence
(time runing left to right),%
\begin{equation}
X_{2}^{1/2}Y_{1}^{1/2}(X_{1}X_{2})^{1/2}Y_{1}^{-1/2}Z_{1}^{1/2},
\end{equation}
which, as required, performs the $U^{\text{\textsc{CNOT}}}$ mapping in the
Heisenberg picture: $\{\sigma_{1}^{z},\sigma_{1}^{x},\sigma_{2}^{z},\sigma
_{2}^{x}\}\rightarrow\{\sigma_{1}^{z},\sigma_{1}^{x}\sigma_{2}^{x},\sigma
_{1}^{z}\sigma_{2}^{z},\sigma_{2}^{x}\}$\cite{Gottesman-Heis}. Though similar
in spirit to decompositions of $U^{\text{\textsc{CNOT}}}$ given elsewhere for
other systems, e.g. \cite{Cirac}, expression (2) is presented in the general
language of Pauli rotation operators, making it applicable to any physical
implementation. It can, with simple algebra, be adapted to systems where the
core two-qubit gate is other than $(X_{1}X_{2})^{1/2}$\cite{Rig/Dev GDQI}.

\begin{figure}[ptb]
\begin{center}
\includegraphics[width=2.3in
]{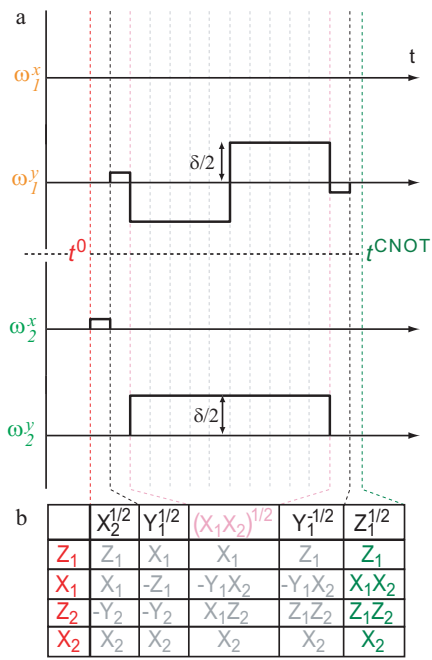}
\end{center}
\caption{Pulse sequence for $U^{\text{\textsc{CNOT}}}$ using FLICFORQ. (a)
Amplitude of in-phase ($\omega^{x}$) and quadrature ($\omega^{y}$) components
of resonant RF signal on first (top) and second (bottom) qubits. Protocol
constructs $U^{\text{\textsc{CNOT}}}$ from sequence of five \textquotedblleft
primitive\textquotedblright\ $\pi/2$ rotations. One-qubit $\sigma^{x}$ and
$\sigma^{y}$ pulses have amplitude $\delta/8$ and duration $t_{2}^{sync}%
=4\pi/\delta$; two-qubit pulses have amplitude $\delta/2$. The $\sigma^{z}$
rotation occurs last, where it can be ignored if followed by measurement in
computational basis. Gate is completed at time $t^{CNOT}$. (b) Description of
sequence in Heisenberg picture\cite{Gottesman-Heis}. First and last columns
(red, green) are connected by $U^{\text{\textsc{CNOT}}}$.}%
\end{figure}

The full $U^{\text{\textsc{CNOT}}}$ pulse sequence is constructed by
concatenating the pulses generating each of the constituent rotations in
expression (2) (see figure 3).

We must briefly comment on how the difference between the Larmor frequencies
is dealt with. In the absence of irradiation the natural evolution of the
system consists of continuous rotations of each qubit about the $\sigma^{z}$
axis, resulting in a time-dependent phase between the qubits that vanishes
every $t_{o}^{sync}=2\pi/\delta$. This phase is unimportant when considering
one-qubit gates, as compensatory $\sigma^{z}$ rotations may be realized
through simple waiting periods\cite{NMR2}. However, it must be taken into
account when doing two-qubit rotations by initiating two-qubit pulses only at
$t_{m}^{sync}=mt_{o}^{sync}$ for integer $m$, i.e. when the qubits are in
synchrony. This condition can be met by using one-qubit pulse amplitudes such
that the one-qubit $\sigma^{x}$ and $\sigma^{y}$ rotations last $t_{m}^{sync}%
$. For the above chosen parameters, an amplitude of $\delta/8$ for $\pi/2$
pulses is convenient (finer rotations may be generated by weaker pulses with
the same duration), corresponding to $t_{2}^{sync}=4\pi/\delta$. The
associated timing grid is shown with dashed vertical lines in figure 3. This
scheme is generalizable to multiqubit registers (see below).

We have simulated the pulse sequence of figure 3 by numerically solving a set
of fifteen coupled differential equations describing each component of the
two-qubit density operator\cite{Rig/Dev TQSO}. The simulation technique is
exact in the sense that it does not rely on any approximations or perturbative
expansions of the time-dependent Hamiltonian. Figure 4 shows the results of a
simulation of two representative evolutions. The simultaneous vanishing of
each component of the two reduced density operators indicates the generation
of a fully entangled two-qubit state\cite{Nielsen and Chuang}.
\begin{figure}[htb]
\begin{center}
\includegraphics[width=3.1in
]{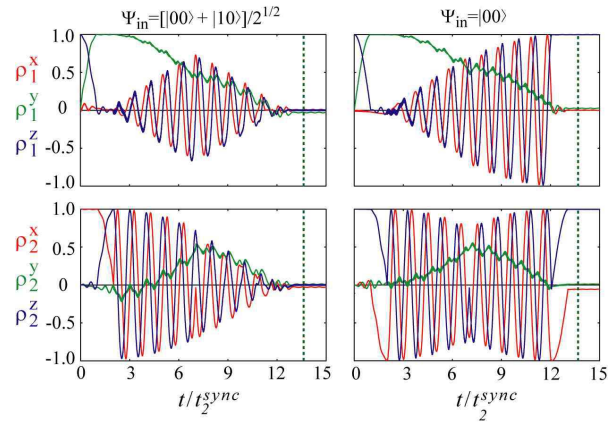}
\end{center}
\caption{Evolution of sample input states during $U^{\text{\textsc{CNOT}}}$
sequence. Components of the reduced density operators $\rho_{1}=Tr_{2}%
\mathbf{\rho}$ (row 1) and $\rho_{2}=Tr_{1}\mathbf{\rho}$ (row 2). $\rho_{1}$
and $\rho_{2}$ are plotted in reference frames rotating at $\omega_{1}^{z}$
and $\omega_{2}^{z}$, resp. Dashed vertical line denotes $t^{CNOT}$. Error
visible at $t^{CNOT}$ is due to Bloch-Siegert shift and effect of coupling
during one-qubit rotations\cite{comment 2}.}%
\end{figure}

What level of gate fidelity can we expect from this scheme? We first discuss
the error in one-qubit gates. Choosing the one-qubit pulse time to be
$t_{2}^{sync}$ means that the coupling term $\omega^{xx}$ will perform a
parasitic rotation in this time by an angle $\arccos(\omega^{xx}t_{2}%
^{sync}).$ This rotation can be nullified altogether using dynamic decoupling
schemes, as is done in NMR\cite{NMR2}. In the present system, this would be
done by performing appropriately timed $\pi$ rotations about $\sigma^{y}$,
which anticommutes with the coupling term $\sigma_{1}^{x}\sigma_{2}^{x}$.
However, for the range of practical circuit parameters $\delta\gtrsim
0.1\omega_{o}^{z}$ and $\omega^{xx}\lesssim0.01\omega_{o}^{z}$, the one-qubit
gate error rate resulting from this parasitic rotation is already below
$10^{-3}$, or two orders of magnitude better than the fidelity of presently
available readout schemes\cite{Dev/Mart}, and the correction is an easily
dispensable luxury. Also, though our simulations have used only square pulses,
realistic pulse shapes should not cause a significant further loss of fidelity
in the one-qubit operations, since, as is commonplace in NMR, pulse shapes
requiring far less bandwidth could be used\cite{NMR2}.

The two-qubit gate error rate will likely be dominated by errors resulting
from imperfect RF pulses. The strength of the qubit-qubit interaction depends
strongly on the amplitude of the simultaneous RF signals, so entangling gates
will be sensitive to jitter or ringing in the pulse amplitudes. This problem
could be minimized by using slowly-rising pulses which require less bandwidth
rather than trying to approximate square pulses. The pulses could be
constructed so that \textit{intended} one-qubit rotations are implemented
during the rise time.

Nonetheless, there is still some error present in our simulations of two-qubit
gates, even though we have used ideal pulses. We have verified that this can
be attributed to the counter-rotating term in the rotating wave
framework\cite{Bloch-Siegert}, as the qubits are irradiated with strong fields
for many Larmor periods during a two-qubit gate. This error can be reduced by
choosing a stronger coupling $\omega^{xx}$, thereby reducing the time required
to generate entanglement, or by reducing the detuning $\delta$, which reduces
the required field strength. Since these changes would increase the one-qubit
gate error rate due to the fixed coupling, an NMR-style decoupling scheme will
likely be needed once we require a gate error rate $\lesssim10^{-3}$.

We believe a main advantage of the gate scheme presented in this paper is that
it can be directly generalized to multiqubit registers with minimal extra
hardware. A fixed linear coupling reactance between all \textit{pairs} of
qubits could easily be achieved by coupling each qubit to a common
superconducting island, loop or cavity. Selective one-qubit gates would be
realized by applying an RF signal at just the target qubit transition
frequency, while the protocol generating \textsc{D} or the universal two-qubit
rotation $(X_{1}X_{2})^{1/2}$ could be realized on any pair by simultaneously
applying RF signals at the resonant frequencies of the two targeted qubits.
Since each qubit in the register would be detuned from all the others, all
these write pulses could be multiplexed on a \textit{single} RF control line,
a decisive advantage in seeking to limit stray couplings to the environment or
crosstalk between qubits. Applying \textsc{D} to several qubits in a pairwise
fashion would allow the direct production of multiqubit entangled states of
the form $|GHZ\rangle=(|0\rangle^{\otimes n}+e^{i\phi}|1\rangle^{\otimes
n})/\sqrt{2}$, which, for $n>2$ can display maximal violations of Bell-type
inequalities\cite{GHZ}.

The authors are grateful to Alexandre Blais, Daniel Esteve, Steve
Girvin, Rob Schoelkopf, Irfan Siddiqi, Cristian Urbina and Denis
Vion for helpful discussions. This work was supported by ARDA (ARO
Grant DAAD19-02-1-0044) and the NSF (Grant DMR-0072022).


\begin{thebibliography}{99}
\bibitem {note on classical comp}Though it leads to the simple pictorial
representation of figure 2, the quantum treatment of the RF fields is not
required, as similar conclusions would be obtained with a classical rotating
wave approximation.
\bibitem {comment 1}More generally, any two Rabi frequencies whose sum equals
$\delta$ could be used to entangle the qubits.
\bibitem {comment 2}The error in each of the five $\pi/2$ rotations in the
\textsc{CNOT}\ sequence is $\simeq10^{-3}$, giving a \textsc{CNOT}\ gate error
rate $\simeq5\times10^{-3}$.
\end{thebibliography}
\end{document}